\documentclass[]{spie}  

\usepackage{amsmath,amsfonts,amssymb}
\usepackage{graphicx}
\usepackage{textcomp}

\usepackage[colorlinks=true, allcolors=blue]{hyperref}

\title{Design of an aluminum nitride based electro-optic phase modulator and photonic switch for next
generation scalable photonic integrated circuits}

\author[a,b,*]{Suat Icli}
\author[a]{Rangana Banerjee Chaudhuri}
\author[a]{Elena Jordan}
\author[a]{Fatemeh Salahshoori}
\author[a,b,c]{\newline Tanja E. Mehlstäubler}
\affil[a]{Physikalisch-Technische Bundesanstalt, D-38116 Braunschweig, Germany}
\affil[b]{Institut für Quantenoptik, Leibniz Universität Hannover, D-30167 Hannover, Germany}
\affil[c]{Laboratorium für Nano- und Quantenengineering, Leibniz Universität Hannover, Schneiderberg 39, 30167 Hannover, Germany}

\authorinfo{*suat.icli.ext@ptb.de}

\begin{document} 
\maketitle

\begin{abstract}
Electro-optic modulators are fundamental components in atomic physics experiments, including trapped-ion systems used in precision metrology and quantum computing. To enable scalable photonic integration, we design and analyze an integrated photonic electro-optic phase modulator and switch at 411 nm for ytterbium (Yb\textsuperscript{+}) ions using aluminum nitride (AlN) waveguides. We employ finite element method (FEM) simulations to optimize optical confinement, RF impedance matching, and electro-optic modulation efficiency. The phase modulator achieves a DC $V_{\pi}L$ of 178 V cm for TE polarization. The photonic switch, designed with a push-pull electrode configuration, demonstrates a $V_{\pi}L$ of 24 V cm, enabling efficient operation at lower voltages. These results highlight AlN as a candidate for ultraviolet photonic integrated circuits, facilitating high-speed optical modulation for trapped-ion applications. 
\end{abstract}


\section{INTRODUCTION}
\label{sec:intro}  

Modulators play a crucial role in atomic, molecular and optical physics experiments \cite{blumenthal2024enabling}. Among these, trapped-ion systems are a prominent example, where the function of modulators vary from laser stabilization to more advanced operations such as qubit manipulation \cite{lee2003atomic} or individual addressing of ions \cite{lim2025design}. A popular ion species is Yb\textsuperscript{+} ions, which are employed in both optical clocks \cite{huntemann2012high} and quantum computing \cite{pino2021demonstration}. The typical wavelengths in these experiments range from ultraviolet (UV) to infrared (IR), with required modulator frequencies spanning from hundreds of MHz to GHz \cite{olmschenk2007manipulation}. To increase the scalability of such systems by replacing free-space or fiber optics with photonic integrated circuits (PICs), AlN is a promising material. In addition to its compatibility with CMOS fabrication techniques, its Pockels and piezoelectric effects, combined with a large transparency window covering wavelengths from UV to IR, make it an important candidate for realizing both active and passive devices that operate at different wavelengths and perform various functions on the same chip \cite{li2021aluminium}. Previously, AlN electro-optic modulators \cite{xiong2012low} were demonstrated for telecom wavelengths. Their bandwidths were limited due to resonant devices used. In this work, we propose to design and demonstrate wide-bandwidth electro-optic devices (phase modulator and photonic switch in particular) at wavelengths close to UV, with the potential to function at bandwidths of tens of GHz.

In this work, we design and analyze an integrated photonic electro-optic phase modulator and switch at 411 nm for $^{171}$Yb\textsuperscript{+} ions. Using finite element method (FEM) simulations, we optimize field distributions, impedance matching, and modulation efficiency to enable scalable photonic integration for trapped-ion applications. Section ~\ref{sec:design} provides an overview of the relevant equations, materials, and simulation methods used for both devices, while Sec.~\ref{sec:results} presents the simulation results and the estimated device performance followed by the conclusion in Sec.~\ref{sec:conclusion}
\section{Design and Simulation}
\label{sec:design}

\begin{figure} [!ht]
\begin{center}
\begin{tabular}{cc}

    \includegraphics[height=4.5cm]{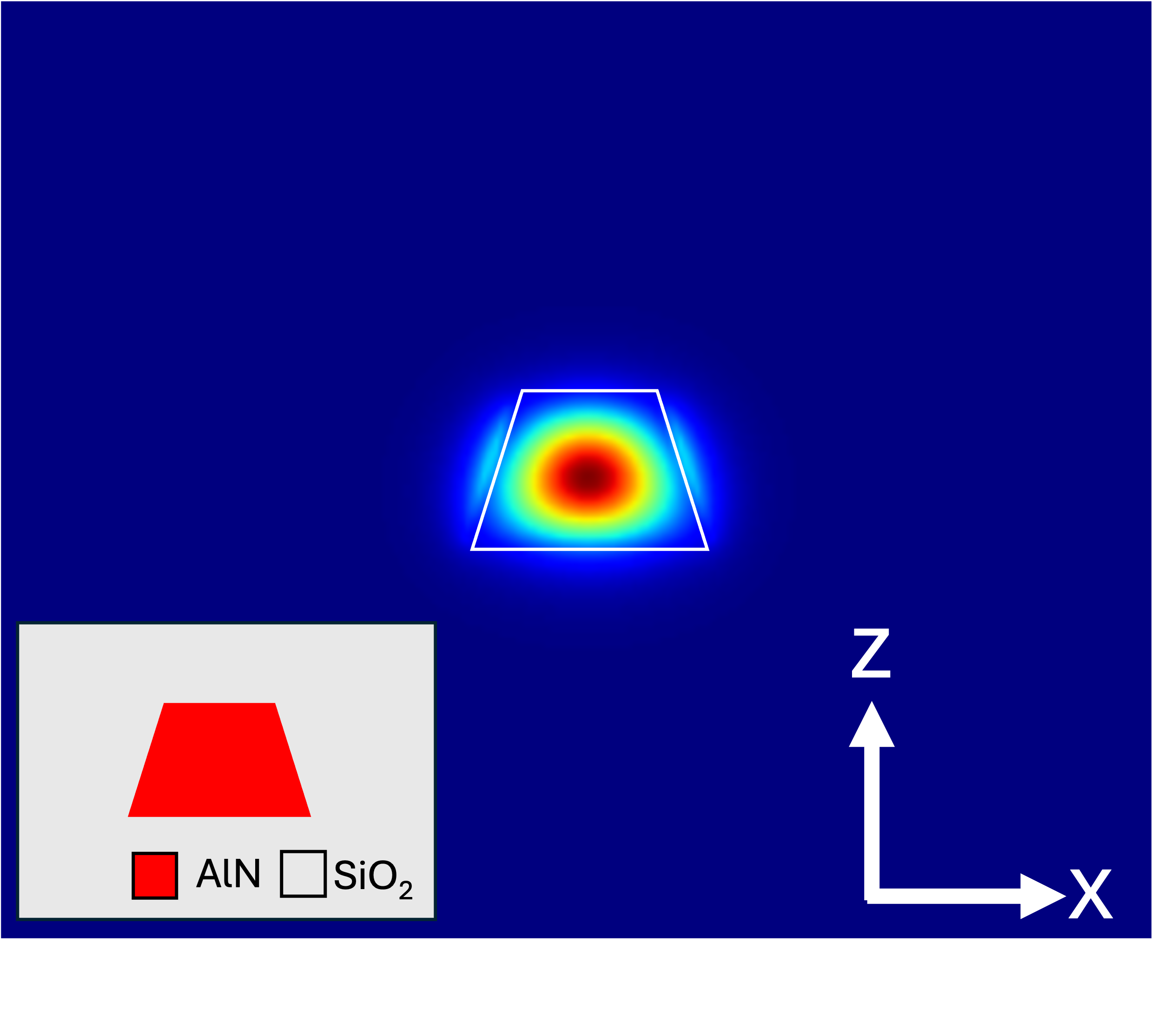} & 
    \includegraphics[height=4.5cm]{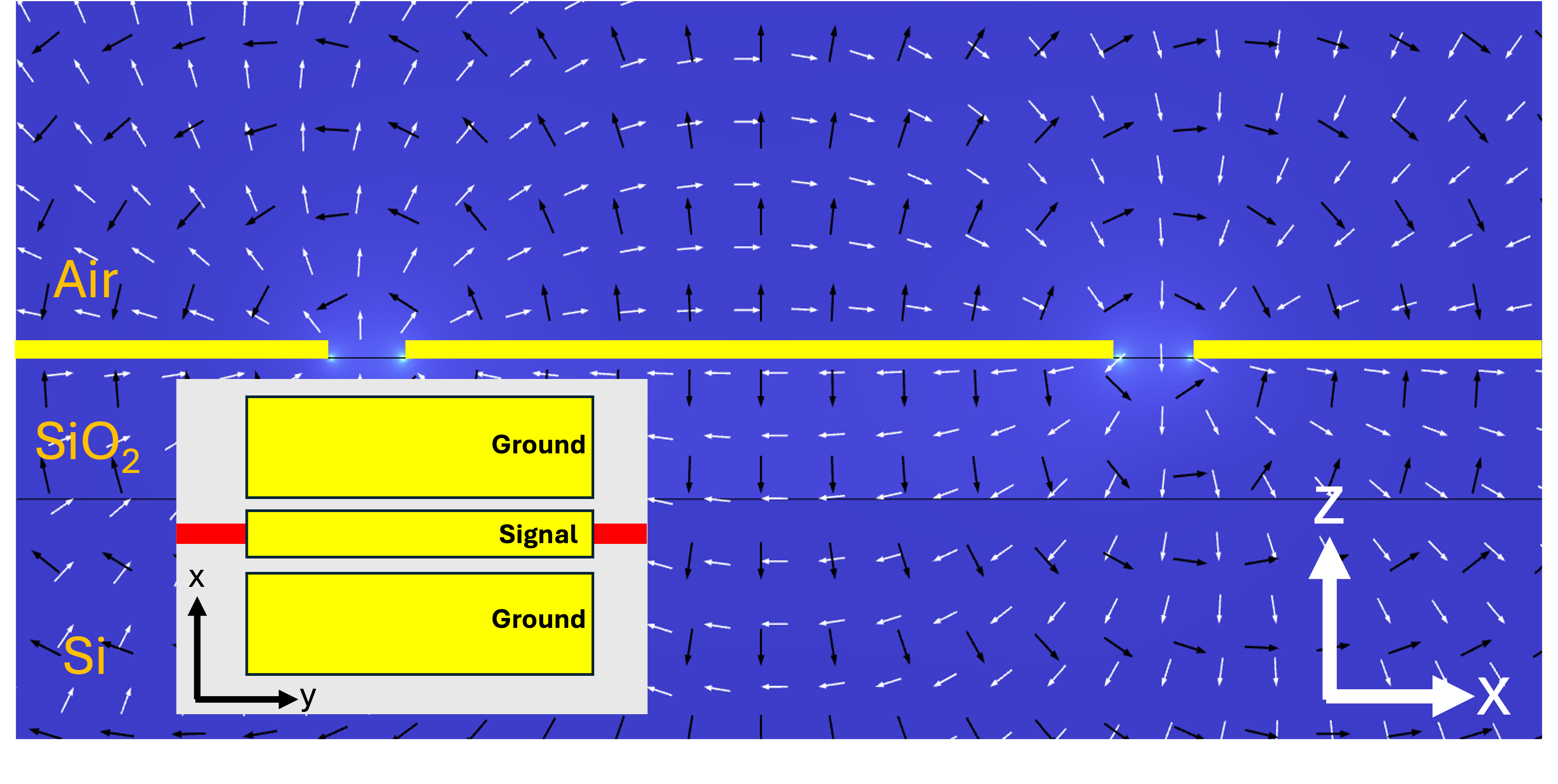} \\ 
    (a) & (b) 
\end{tabular}
\end{center}
\caption[example] 
{ \label{fig:fields} 
(a) Simulated optical mode field with the inset showing the device cross-section. (b) Electric and magnetic field distribution at 40 GHz, where arrows represent normalized field directions: white arrows indicate magnetic fields, and black arrows indicate electric fields. The inset shows the top view of device.}
\end{figure}

For c-axis grown AlN, the relevant electro-optic coefficients are $r_{13}$, $r_{33}$, and $r_{51}$. While $r_{13}$ and $r_{33}$ are approximately 1 pm/V, $r_{51}$ is reported to be smaller \cite{xiong2012low}. The refractive index change induced by the applied electric field for TE polarization ($n_x$) and TM polarization ($n_z$) is given by Eq. (\ref{eq:index_change}), where the $E_z$ is the electric field component along the z-axis.
\begin{equation}
    \label{eq:index_change}
    \begin{split}
            &n_{x} = n_{x} - \frac{1}{2}n_{x}^3r_{13}E_{z}\\
            &n_{z} = n_{z} - \frac{1}{2}n_{z}^3r_{33}E_{z} 
    \end{split}
\end{equation}
For efficient modulation, the applied electric field must be maximized within the waveguide. One approach is using electrodes positioned above and below the waveguide to generate a strong vertical field, but this configuration presents fabrication challenges due to complex multilayer processing. Instead, a coplanar ground-signal-ground (GSG) electrode configuration is employed, where the signal electrode is placed directly on top of the waveguide and is flanked by two ground electrodes. This setup provides a balance between maximizing $E_z$ and fabrication feasibility. The designed devices consist of gold electrodes, AlN waveguides with SiO\textsubscript{2} cladding, and a silicon substrate. The cross-section of the waveguide is shown in the inset of Fig.~\ref{fig:fields}a, while the gold electrodes are shown in Fig.~\ref{fig:fields}b. 
\subsection{Optical Simulations}
To optimize the optical design, Ansys Lumerical MODE\textsuperscript{\textregistered} is used. For the optimization, we compute the optical mode properties, including the fundamental mode profile, effective index, and electrode induced losses. The optical simulations focus on maximizing light confinement within the waveguide while minimizing absorption losses from metal electrodes. A key design parameter is the spacing between the waveguide and the electrodes, as reducing this distance enhances electro-optic interaction but can lead to increased optical losses. The optimal spacing is chosen to balance these competing factors.
\subsection{Electrical Simulations}
The electrostatic and RF field distributions are modeled using COMSOL Multiphysics\textsuperscript{\textregistered}. Electrostatic simulations evaluate the impact of electrode placement, dielectric properties, and applied voltage on the electric field distribution. The RF simulations analyze key parameters such as characteristic impedance, propagation constant, and scattering parameters (S-parameters) to ensure efficient RF power transfer. Achieving proper impedance matching is crucial to minimize signal reflections and maximize modulation efficiency. A target impedance of 50 $\Omega$ is chosen to align with standard RF driving circuits. The simulated S-parameters provide insight into the reflection and transmission characteristics of the electrodes, ensuring that the modulator can operate effectively across a wide frequency range.
\subsection{System-level Integration}
To integrate optical and electrical simulation results, Ansys Lumerical INTERCONNECT\textsuperscript{\textregistered} is used for system-level modeling. This tool incorporates the optical and electrical simulation results into a circuit-level model, allowing the evaluation of device performance under realistic operating conditions. The modulated optical spectrum is obtained to study the frequency-dependent modulation response, accounting for impedance mismatch, velocity mismatch, and RF loss.

\section{Results and discussion}
\label{sec:results}
Figure ~\ref{fig:fields}a shows the simulated optical mode profile of the fundamental TE mode at 411 nm, which has a group index of 2.6. The inset shows the materials composing the waveguide region. Additionally, the distance between the signal electrode and the waveguide is optimized to balance multiple factors: minimizing optical losses due to metal absorption, enhancing the $E_z$ field intensity, and maintaining a practical fabrication process. This design approach offers a good compromise between device performance and ease of fabrication.

\subsection{Phase Modulator}
\label{sec:phasemod}
The phase modulator consists of an AlN waveguide and GSG electrodes with a length of 5 mm.
To achieve efficient and wide-bandwidth modulation of optical signals, several key criteria must be fulfilled: ensuring impedance matching between the source, transmission line, and load (typically 50 $\Omega$), minimizing RF propagation loss, and maintaining velocity matching between the optical and RF fields. 
\begin{figure} [htb]
\begin{center}
\begin{tabular}{cc}
\includegraphics[height=4cm]{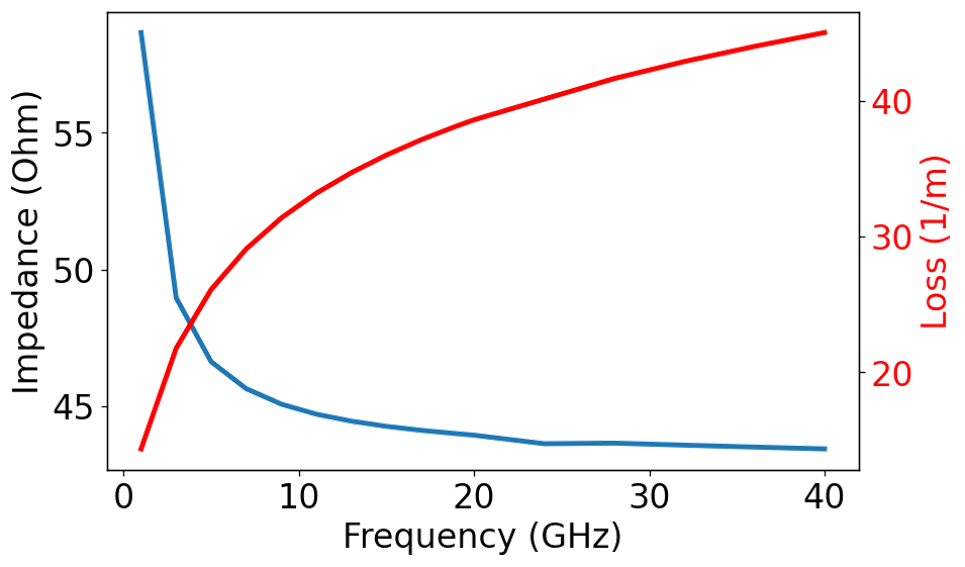} & 
\includegraphics[height=4cm]{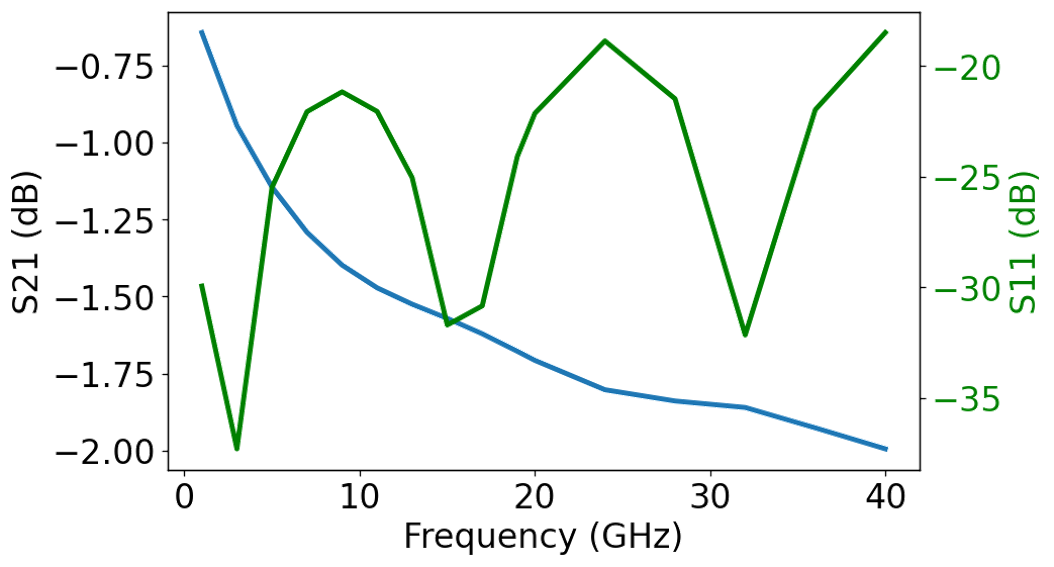} \\
(a) & (b) \\
\multicolumn{2}{c}{\includegraphics[height=4cm]{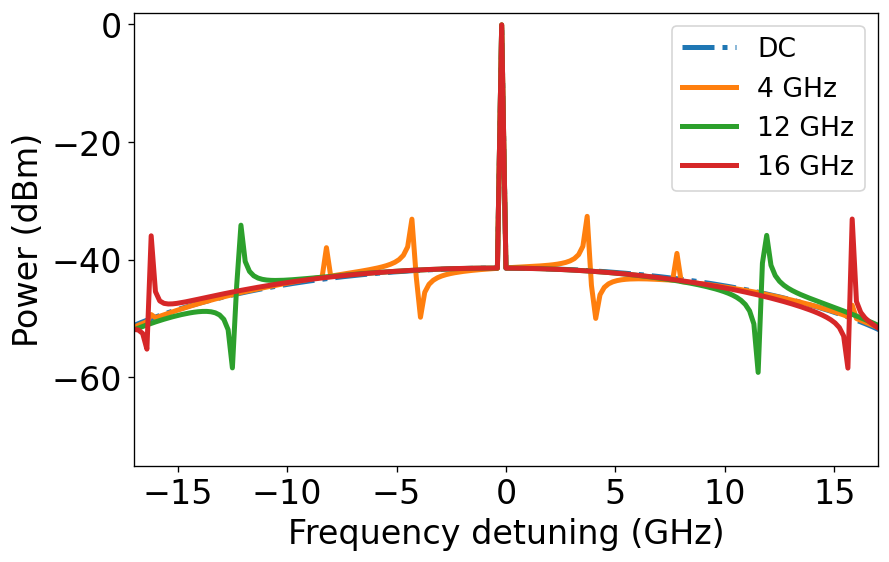}} \\
\multicolumn{2}{c}{(c)}
\end{tabular}
\end{center}
\caption[example] 
{ \label{fig:phasemod} 
(a) Impedance and attenuation coefficient as a function of frequency. (b) Electrical $S_{11}$ and $S_{21}$ parameters. (c) Modulated optical spectrum at 11 V.}
\end{figure}

Figure~\ref{fig:fields}b presents the simulated electric and magnetic field distributions at 40 GHz, where white arrows indicate magnetic fields and black arrows represent electric fields. The RF index at this frequency is 1.86, and further velocity matching can be achieved using capacitively loaded electrodes \cite{shin2010conductor}. Other key parameters such as characteristic impedance, propagation constant, and S-parameters can be extracted from the simulations. Figure~\ref{fig:phasemod}a illustrates the impedance and attenuation characteristics as a function of frequency, showing that as impedance decreases, losses increase with frequency.
Additionally, S-parameters, which describe the transmitted and reflected power ratios as a function of frequency, are presented in Fig.~\ref{fig:phasemod}b. The transmission coefficient, $S_{21}$, is greater than -2 dB, while the reflection coefficient, $S_{11}$, remains below -18 dB up to 40 GHz, demonstrating efficient signal delivery with minimal reflection. Optical eigenmode simulations are then performed to calculate the refractive index change based on Eq. (\ref{eq:index_change}), which is used to determine the voltage-length product, $V_{\pi}L$. At DC, $V_{\pi}L$ is 178 V cm and 197 V cm for the TE and TM modes, respectively. Since RF and DC $V_{\pi}L$ are related by $V_{\pi,RF}=V_{\pi,DC}\times10^{-EO S_{21}/20}$ \cite{zhang2022systematic}, a higher $V_{\pi}L$ is expected at higher frequencies due to increased impedance mismatch and loss. EO $S_{21}$ represents the electro-optic S-parameter. To assess the effects of impedance mismatch, velocity mismatch, and RF loss, the RF and optical simulation results can be combined to simulate the modulated optical spectrum using Lumerical INTERCONNECT. These simulations provide a comprehensive understanding of how electrical and optical properties interact, helping to identify potential performance bottlenecks. The resulting modulated spectrum at 11 V for TE polarization is plotted in Fig.~\ref{fig:phasemod}d, where the modulation depth and spectral characteristics offer insight into the device’s operational bandwidth and efficiency. The observed low modulation depth can be attributed to the high $V_{\pi}L$ product, which can be reduced by narrowing the signal electrode width. While this approach enhances the $E_z$ field and increases the RF index, it also leads to higher RF losses and impedance variations, ultimately limiting modulation efficiency.

\subsection{2x2 Electro-optic Switch}
\label{sec:lightswitch}

\begin{figure} [htb]
\begin{center}
\begin{tabular}{cc}

\includegraphics[height=4cm]{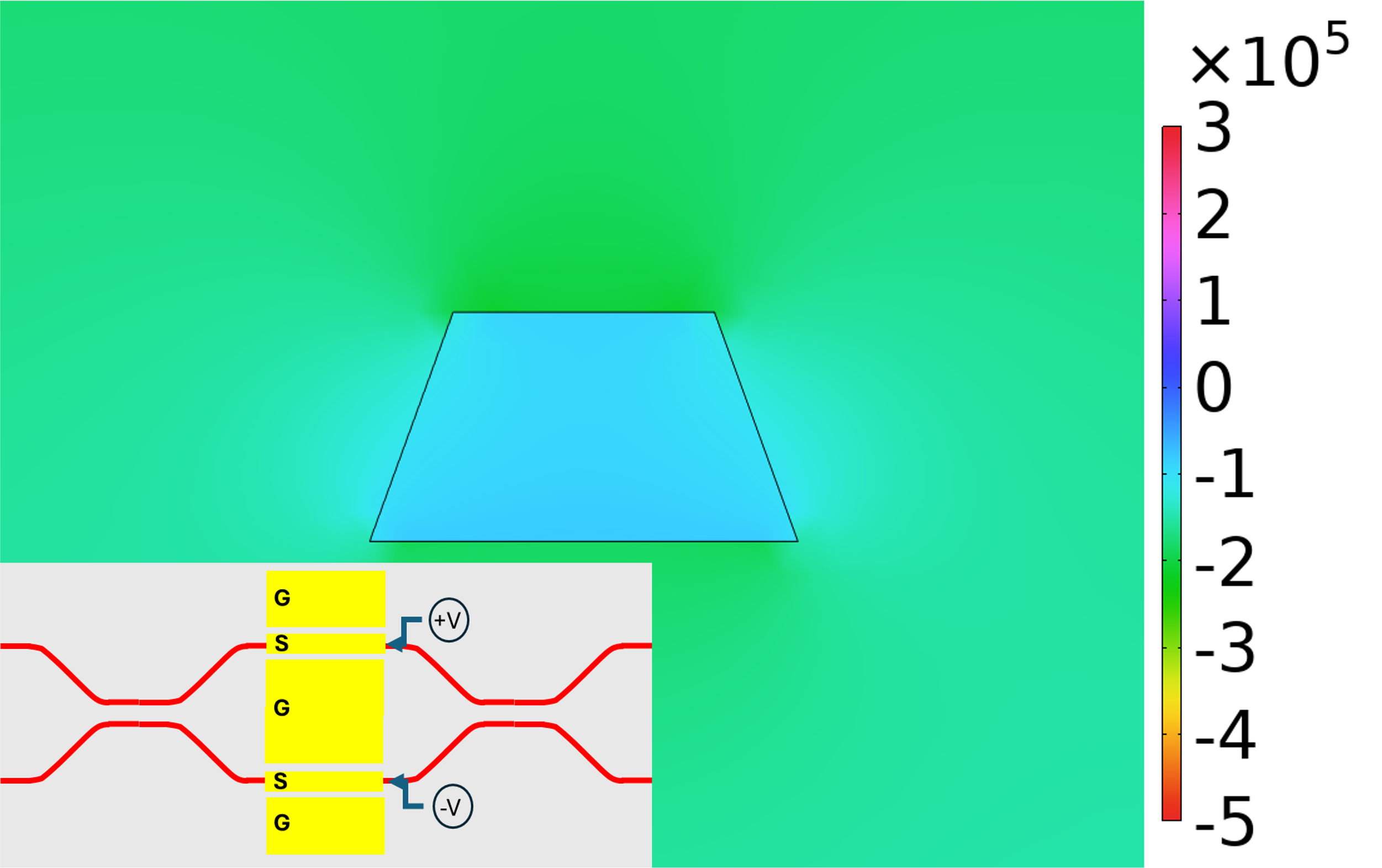} & 
\includegraphics[height=4cm]{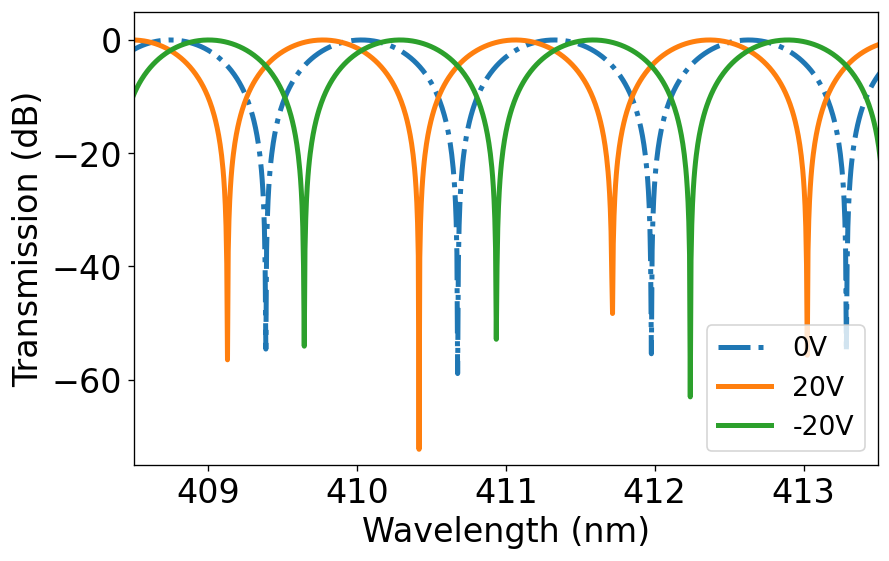} \\
(a) & (b) \\
\includegraphics[height=4cm]{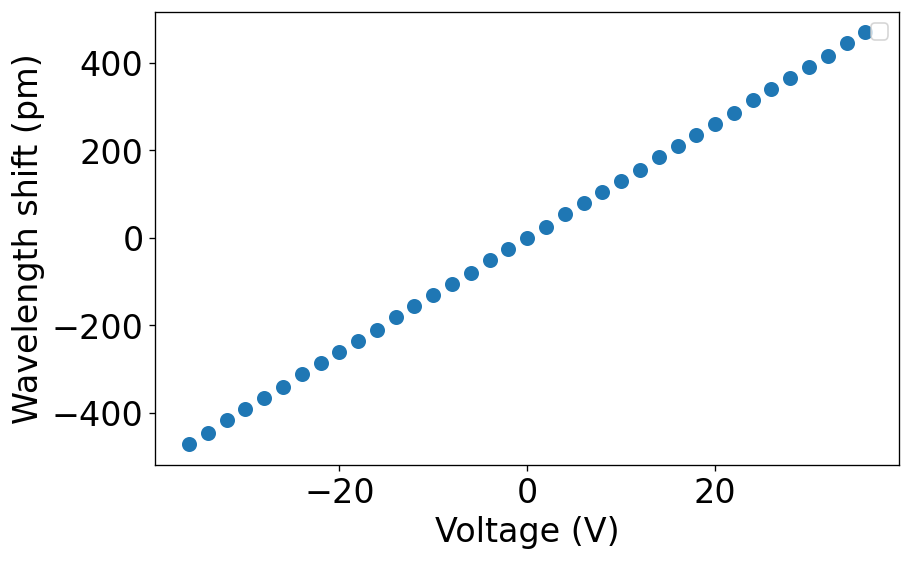} & 
\includegraphics[height=4cm]{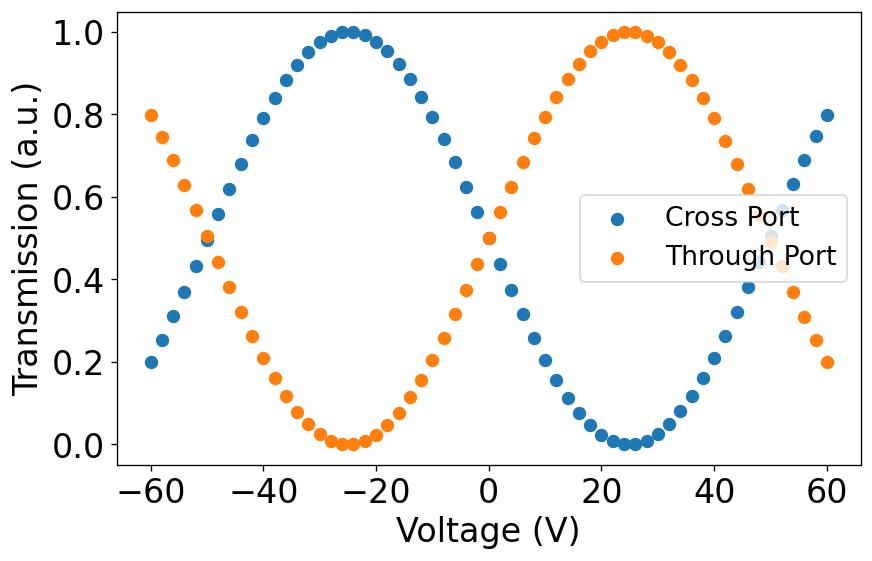} \\
(c) & (d)
\end{tabular}
\end{center}
\caption[example] 
{ \label{fig:switch} 
(a) Optical mode field distribution at 1V, with the color bar in units of V/m. (b) Transmission spectrum at the through port. (c) Wavelength shift as a function of applied voltage. (d) Transmission at the through and cross ports for different applied voltages.}
\end{figure}

The electro-optic switch is based on a Mach-Zehnder interferometer (MZI) with an unbalanced arm length to operate around the quadrature point, ensuring a linear response and reduced amount of the required driving voltage.  To achieve a push-pull configuration, electrodes are arranged in a ground-signal-ground-signal-ground (GSGSG)  configuration, where opposite polarities are applied to the signal electrodes on each arm of the MZI. This configuration improves the switching efficiency by reducing the required drive voltage.  The electrodes have a length of 5 mm. The DC $E_z$ field distribution at 1V is plotted in Fig.~\ref{fig:switch}a, with the inset showing the device layout.  The average field inside the waveguide is determined to be $|E_z|=93$ kV/m.
The optical spectrum at the through port of the switch for different bias voltages is plotted in Fig.~\ref{fig:switch}b, revealing a free spectral range of 1.30 nm around 411 nm. 

By analyzing the transmission dips, the wavelength shift as a function of applied voltage is extracted as 13.05 pm/V, as shown in Fig.~\ref{fig:switch}c. This confirms the expected linear response to the applied field, a characteristic feature of electro-optic modulation in the device. The switching characteristics at 411 nm are plotted in Fig.~\ref{fig:switch}d, where the transmission at the through and cross ports is analyzed as a function of voltage. The result reveals a $V_{\pi}L$ of 24 V cm for TE polarization, demonstrating efficient optical switching at reduced drive voltages. The observed switching performance could be enhanced through further optimization of the electrode design and waveguide confinement, leading to improved switching contrast while reducing the required drive voltage.

\section{Conclusion}
\label{sec:conclusion}
We designed, simulated and analyzed integrated photonic electro-optic modulators and switches at 411 nm for trapped-ion applications. Using FEM simulations, we optimized impedance matching and field distributions within the waveguide. The phase modulator balanced fabrication feasibility with modulation efficiency, while the 2×2 electro-optic switch demonstrated efficient push-pull operation with a $V_{\pi}L$ of 24 V cm. These results highlight AlN as a promising material for UV photonic integrated circuits. Future work could focus on reducing RF $V_{\pi}L$ and enhancing light-matter interaction using different structures.

\acknowledgments 
This project has received funding from the European Union's Horizon Europe research and innovation programme under Grant Agreement No \#101135845. We acknowledge funding from Germany’s Excellence Strategy within the Cluster of Excellence PhoenixD (EXC-2122). The authors would like to thank University of Twente for providing the refractive index data for AlN.
\bibliography{report} 
\bibliographystyle{spiebib} 

\end{document}